# Trompe L'oeil Ferromagnetism: magnetic point group analysis

Sang-Wook Cheong* and Fei-Ting Huang

Rutgers Center for Emergent Materials and Department of Physics and Astronomy, Rutgers University

*Corresponding author: sangc@physics.rutgers.edu

**ABSTRACT**

**Ferromagnetism can be characterized by various unique phenomena such as non-zero magnetization (inducing magnetic attraction/repulsion), diagonal piezomagnetism, nonreciprocal circular dichroism (such as Faraday effect), odd-order (including linear) anomalous Hall effect, and magneto-optical Kerr effect. We identify all broken symmetries requiring each of the above phenomena, and also the relevant magnetic point groups (MPGs) with those broken symmetries. All of ferromagnetic point groups, relevant for ferromagnets, ferri-magnets and weak ferromagnets, can certainly exhibit all of these phenomena, including non-zero magnetization. Some of true antiferromagnets, which are defined as magnets with MPGs that do not belong to ferromagnetic point groups, can display these phenomena through magnetization induced by external perturbations such as applied current, electric fields, light illumination, and strain. Such MPGs are identified for each external perturbation. Since high-density and ultrafast spintronic technologies can be enabled by antiferromagnets, our findings will be an essential guidance for the future magnetism-related science as well as technology.**

## INTRODUCTION

Magnetic states exhibiting non-zero net magnetization include ferromagnets, ferrimagnets or weak ferromagnets (referred to canted antiferromagnets). Due to the non-zero magnetization, those magnetic states exhibit unique physical phenomena such as magnetic attraction, various magneto-optical properties (magneto-optical Kerr (MOKE), Faraday, and magnetic circular dichroism) and anomalous Hall-type effects[1-12]. The anomalous Hall-type effects include anomalous Hall, anomalous Ettingshausen, anomalous Nernst, or anomalous thermal Hall effects[7,12-17]. It turns out that these phenomena that traditionally thought to occur only in magnetic states with non-zero magnetization can take place in certain antiferromagnets with seemingly zero net magnetization, sometime in the presence of external perturbations such as stress or electric fields or with time evolution. These cases have been called Trompe L'oeil Ferromagnetism[18].

It turns out that broken symmetries can be associated with new order parameters and emergent phenomena. Herein, we identify the exact broken symmetries associated with each of these phenomena: non-zero magnetization, diagonal piezomagnetism[19], circular dichroism[2, 6, 20, 21], nonreciprocal circular dichroism (such as Faraday effect), odd-order (including linear) anomalous Hall effect (AHE), and MOKE[22]. For these analyses, we utilize the concept of Symmetry Operational Similarity (SOS), and also classify the corresponding magnetic point groups (MPGs) with those broken symmetries, i.e. each of the above phenomena.

Observable physical phenomena can occur or non-zero measurable can be detected when specimens have SOS with the combination of measurable (or experimental setup for measurable) and specimen environments (such as applied external stress, electric fields or magnetic fields) or specimens with specimen environments have SOS with measurable. This SOS relationship includes when specimens have more, but not less, broken symmetries than the combination of measurable and specimen environments or specimens with specimen environments have more, but not less, broken symmetries than measurable. In other words, in order to have a SOS relationship, specimens "cannot have higher symmetries" than the combination of measurable and specimen environments or specimens with specimen environments cannot "have higher symmetries" than measurable. The power of the SOS approach lies in providing simple and physically transparent views of otherwise unintuitive phenomena in complex materials, without considering specific coupling terms or the relevant Hamiltonians. Furthermore, this approach can be leveraged to identify new materials that exhibit potentially desired properties as well as new phenomena in known materials.

To find the requirements of broken symmetries for various phenomena, we, first, define the general symmetry operation notations for three orthogonal *x*, *y*, and *z* axes such as $\mathbf{R_x}$=2-fold rotation around the *x* axis, $\mathbf{M_x}$=mirror reflection with mirror perpendicular to the *x* axis, $\mathbf{I}$=space inversion, $\mathbf{T}$=time reversal, etc. Then, we have these general relationships: $\mathbf{R_x} \otimes \mathbf{R_y} = \mathbf{R_z}$, $\mathbf{M_z} \otimes \mathbf{R_y} = \mathbf{M_x}$, $\mathbf{M_x} \otimes \mathbf{R_z} = \mathbf{M_y}$, $\mathbf{M_x} \otimes \mathbf{R_y} = \mathbf{M_z}$, $\mathbf{M_y} \otimes \mathbf{M_z} = \mathbf{R_x}$, $\mathbf{M_x} \otimes \mathbf{M_z} = \mathbf{R_y}$, $\mathbf{M_x} \otimes \mathbf{M_y} = \mathbf{R_z}$, $\mathbf{M_x} \otimes \mathbf{R_x} = \mathbf{M_y} \otimes \mathbf{R_y} = \mathbf{M_z} \otimes \mathbf{R_z} = \mathbf{I}$, $\mathbf{M_x} = \mathbf{R_x} \otimes \mathbf{I}$, $\mathbf{M_y} = \mathbf{I} \otimes \mathbf{R_y}$, and $\mathbf{M_z} = \mathbf{I} \otimes \mathbf{R_z}$ (all of these are commutative.). All of our measurables such as magnetization or optical activity setups have translational symmetry, so we can ignore or freely allow any translations (i.e. any translations are considered as a unit operation). Similarly, when we consider one-dimensional (1D) measurables invariant under any rotations along the 1D direction, then we ignore or freely allow any rotations around the axis (i.e. any rotations around the 1D direction are considered as a unit operation). For example, magnetization along $z$ should be invariant under any rotations along $z$, so we ignore or freely allow any rotations around $z$.[23, 24]

## SYMMETRY OF FERROMAGNETISM

Magnetization ($M$) along $z$, which is an 1D object along $z$, has broken $\{T,M_x,M_y,R_x,R_y,C_{3x},C_{3y}\}$ and unbroken $\{1,I,M_z,R_z\}$, and thus has broken $\{I \otimes T, T, M_x, M_y, R_x, R_y, C_{3x}, C_{3y}\}$ with free rotation along $z$. As discussed above, when we consider the symmetry of 1D objects such as $M$, we always allow any free rotations along the 1D direction. For example, $\bar{4}'$ has unbroken $C_{4z} \otimes I \otimes T$ and broken $I \otimes T$, but when we consider the SOS relationship of $\bar{4}'$ with an 1D object along $z$, then $\bar{4}'$ has unbroken $I \otimes T$ with free rotation along $z$, so $\bar{4}'$ does not have SOS with $M_z$. All and also only ferromagnetic point groups do have broken $\{I \otimes T, T, M_x, M_y, R_x, R_y, C_{3x}, C_{3y}\}$ with free rotation along $z$ or the relevant requirements along $x$ or $y$. The thirty one (31) ferromagnetic point groups include 1, $\bar{1}$, 2, 2′, m, $m'$, 2/m, 2′/$m'$, 2′2′2, $m'm'm$, $m'm'2$, $m'm2'$, 4, $\bar{4}$, 4/m, 42′2′, 4$m'm'$, $\bar{4}2'm'$, 4/$mm'm'$, 3, $\bar{3}$, 32′, 3$m'$, $\bar{3}m'$, 6, $\bar{6}$, 6/m, 62′2′, 6$m'm'$, $\bar{6}m'2'$, 6/$mm'm'$. Note that $\bar{4}'$ does not belong to ferromagnetic point groups.

The presence of non-zero net magnetization in magnetic states in ferromagnetic point groups is sometimes evident, but it is not always. For example, magnetic states depicted in Fig. 1(a)-(d) appear to be antiferromagnetic states with 120° spins without any net magnetic moments. However, all belong to ferromagnetic point groups; Fig. 1(a); $mm'm'$, which has been observed in $Mn_3Sn$ (*Cmc'm'*),[9] Fig. 1(b); $m'mm'$, which has been observed in $Mn_3(Ge,Ga)$ (*Cm'cm'*), Fig. 1(c); 2/$m$; kagome lattice with lattice distortions, shown with solid-line bonds (*P2_1/n*)[25], and Fig. 1(d); $\bar{3}m'$, which has been observed in $Mn_3(Rh,Ir,Pt)$ (*R$\bar{3}m'$*).[7, 26] Indeed, small but non-zero magnetization have been observed, at least, in $Mn_3Sn$ and $Mn_3(Ge,Ga)$.[14, 22]

## SYMMETRY OF THE ODD-ORDER AHE MEASUREMENTS

The Hall effect, a hallmark of how Maxell's equations work in materials, was discovered by Edwin Hall while he was working on his doctoral degree in 1879, and has been well utilized to measure carrier density as well as detect small magnetic fields[27]. This so-called ordinary Hall effect contrasts with the anomalous Hall effect (AHE) in "ferromagnets", which is sometime called extraordinary Hall effect or spontaneous Hall effect[17]. This AHE exists in zero applied magnetic field, and varies linearly with applied electric current, so its sign changes when the current direction is reversed. It was proposed that AHE can exist in "truly antiferromagnetic" systems such as $Mn_3$(Rh,Ir,Pt) with Kagome lattice[7, 26], originating from the Berry curvature. In fact, $Mn_3$(Sn,Ge), forming in the same crystallographic structure with that of $Mn_3$(Rh,Ir,Pt), is experimentally reported to exhibit a significant AHE[9, 13]. However, it turns out that $Mn_3$(Sn,Ge) does exhibit a small, but finite net magnetic moment[14, 22], and the exact experimental situation of $Mn_3$(Rh,Ir,Pt) is presently unclear, partially due to the presence of competing multiple magnetic states in the system, and also the absence of bulk crystal study. Topological Hall effects in skyrmion systems have been reported [28, 29], and occur typically in the presence of external magnetic fields. Note that AHE of antiferromagnets with zero or small magnetization can be particularly useful for the fast sensing of magnetic fields due to the intrinsic fast dynamics of antiferromagnets[30].

Herein, we define AHE as "Transverse voltage induced by applied current in zero magnetic field". The sets of (electric current, +/−), (electric current, h/c), (thermal current, +/ −), and (thermal current, h/c) correspond to the Hall, Ettingshausen, Nernst, and thermal Hall effects, all of which we call Hall-type effects, respectively (+/- means an induced voltage difference and h/c (hot/cold) means an induced thermal gradient, accumulated on off-diagonal surfaces). In terms of symmetry, there is little difference among these four types of Hall effects, so, for example, the existence of non-zero AHE means the presence of non-zero anomalous Nernst effect. With these multi-faceted nature of Hall-type effects, it is imperative to find the accurate relationship among all different kinds of Hall-type effects, and also the requirements to have non-zero values of various Hall-type effects.

From our SOS analysis, we can tell a certain phenomenon is a zero, odd-order, or even-order effect. Herein, we will discuss the requirement of broken symmetries for odd-order anomalous Hall effects. We can have these transformations for the experimental setup for AHE measurements in Fig. 1(e)-(h): (e) ↔ (e) by $M_z$, (e) ↔ (h) by $I$ and $R_z$, (e) ↔ (g) by $T$, $M_x$ and $R_y$, and (e) ↔ (f) by $M_y$ and $R_x$. Thus, Odd-order AHE$_{yx}$ means Odd-order AHE with current along $x$ and Hall voltage along $y$ in Fig. 1(a), and the experimental setup to measure odd-order AHE$_{yx}$ has unbroken {$1,I,M_z,R_z$} and broken {$T,M_x,M_y,R_x,R_y,C_{3x}$}. Now, any specimens having SOS with the above experimental setup will show Odd-order AHE$_{yx}$ when they have "unbroken {**1,I,M$_z$,R$_z$**}⊗broken {**T,M$_x$,M$_y$,R$_x$,R$_y$,C$_{3x}$**}" = broken {**I⊗T,T,M$_x$,M$_y$,R$_x$,R$_y$,C$_{3x}$,M$_z$⊗T,R$_z$⊗T**}. The rest independent ones can be either broken or unbroken. For example, for broken {**I**}, basically, Odd-order AHE of the original domain is the same with that of the domain after space inversion. Emphasize that the requirements for anomalous Ettingshausen, anomalous Nernst, and anomalous thermal Hall effects are identical to those for AHE. It turns out that all ferromagnetic point groups[31, 32] can have non-zero net magnetic moments, and do have broken {**I⊗T,T,M$_x$,M$_y$,R$_x$,R$_y$,C$_{3x}$,M$_z$⊗T,R$_z$⊗T**} with free rotation along $z$ when the net magnetic moments are along $z$, broken {**I⊗T,T,M$_y$,M$_z$,R$_y$,R$_z$,C$_{3z}$,M$_x$⊗T,R$_x$⊗T**} with free rotation along $x$ when the net magnetic moments are along $x$, and broken {**I⊗T,T,M$_x$,M$_z$,R$_x$,R$_z$,C$_{3x}$,M$_y$⊗T,R$_y$⊗T**} with free rotation along $y$ when the net magnetic moments are along $y$ – the relevant MPGs are listed in Figs. 2-4. As discussed earlier, all magnetic states in Fig. 1(a)-(d) belong to ferromagnetic point groups, so do exhibit Odd-order AHE. Since all magnetic states in Fig. 1(a)-(d) do accompany non-zero magnetic moments, the AHEs are, in fact, linear with applied electric current. MPGs for Odd-order AHE$_{yx}$, requiring broken {**I⊗T,T,M$_x$,M$_y$,R$_x$,R$_y$,C$_{3x}$,M$_z$⊗T,R$_z$⊗T**}, and those for Odd-order AHE$_{yx,xy}$, requiring broken {**I⊗T,T,M$_{xy}$,M$_{yx}$,R$_{xy}$,R$_{yx}$,C$_{3xy}$,M$_z$⊗T,R$_z$⊗T**}, are listed in Fig. 2. MPGs for Odd-order AHE$_{zy}$, requiring broken {**I⊗T,T,M$_y$,M$_z$,R$_y$,R$_z$,C$_{3y}$,M$_x$⊗T,R$_x$⊗T**}, and those for Odd-order AHE$_{zx}$, requiring broken {**I⊗T,T,M$_x$,M$_z$,R$_x$,R$_z$,C$_{3x}$,M$_y$⊗T,R$_y$⊗T**}, are listed in Fig. 3 and 4, respectively. In all ferromagnetic point groups that can have non-zero net moments along $z$ ($x$ or $y$), Odd-order AHE$_{yx}$ (AHE$_{zy}$ or AHE$_{zx}$) is same with Odd-order AHE$_{xy}$ (AHE$_{yz}$ or AHE$_{xz}$) except their sign difference (i.e. they are anti-symmetric). However, there are three types of non-ferromagnetic point groups which allow Odd-order AHE; (1) MPGs of C$_{3z}$ can have Odd-order AHE$_{zx}$ or AHE$_{zy}$ with broken {**I⊗T,T,M$_z$,M$_y$,R$_z$,R$_y$,C$_{3y}$,M$_x$⊗T,R$_x$⊗T**}, but they have zero Odd-order AHE$_{xz}$ or AHE$_{yz}$ due to unbroken C$_{3z}$: the examples are 3, $\bar{3}$, 32, 3$m$, $\bar{3}m$, 6′, $\bar{6}′$, 6′/$m$′, 6′22′, 6′$mm$′, $\bar{6}′m$2′, 6′/$m$′$mm$′ for Odd-order AHE$_{zy}$ and 3, $\bar{3}$, 32′, 3$m$′, $\bar{3}m$′, 6′, $\bar{6}′$, 6′/$m$′, $\bar{6}′m$′2 for Odd-order AHE$_{zx}$. (2) MPGs of C$_{4z}$⊗T or C$_{4z}$⊗I⊗T can have Odd-order AHE$_{yx}$ and Odd-order AHE$_{xy}$ with broken {**I⊗T,T,M$_x$,M$_y$,R$_x$,R$_y$,C$_{3x}$,C$_{3y}$,M$_z$⊗T,R$_z$⊗T**}: the examples are 4′, $\bar{4}$′, 4′/$m$, 4′2′2, 4′$m$′$m$, $\bar{4}$′2′$m$, $\bar{4}$′$m$′2, 4′/$mm$′$m$. These MPGs are truly antiferromagnetic without any net magnetic moment, but can exhibit odd-order AHE. All of these MPGs do have unbroken 4′ or $\bar{4}$′, so the odd-order AHEs in these point groups are associated with symmetric tensors, unlike antisymmetric tensors for odd-order AHEs in ferromagnetic space groups. For example, for $\bar{4}'$ point group, $(+J_x,+E_y)$ becomes $(+J_y,+E_x)$ under $\bar{4}'$, while the point group is invariant, so the relevant conductivity tensor components are symmetric. These Odd-order AHEs in truly antiferromagnetic states occur without non-zero magnetic moment, so must be high-order effects. Odd-order AHEs with symmetric tensors have never been reported and will be an exciting new research direction. (3) Cubic MPGs allow Odd-order AHE$_{yx,xy}$ as well as Odd-order AHE$_{xy,yx}$: 23, $m\bar{3}$, 4'32', $\bar{4}'3m'$, $m\bar{3}m'$.

It turns out that this Odd-order AHE corresponds to Off-diagonal even-order current-induced magnetization. When we consider the experimental setup for measuring magnetization along z induced in an even-order by current along *x*, we can readily find out that the relevant requirement for non-zero Off-diagonal even-order current-induced magnetization is broken {$I \otimes T, T, M_x, M_y, R_x, R_y, C_{3x}, M_z \otimes T, R_z \otimes T$}, which is identical with the requirement for Odd-order AHE$_{yx}$. Thus, we can conclude that Odd-order AHE results from Off-diagonal even-order current-induced magnetization. Zeroth-order current-induced magnetization is considered the cause of Linear AHE in ferromagnetic point groups with non-zero net magnetization in the presence of no current.

We emphasize that our SOS approach can tell if a certain phenomenon is zero, non-zero odd-order, or non-zero even-order effect, and broken {$I \otimes T, T, M_x, M_y, R_x, R_y, C_{3x}, M_z \otimes T, R_z \otimes T$} is, in fact, the requirement for Odd-order AHE. Recently, the concept of altermagnetism was introduced: their ordered spins are truly antiferromagnetic, but can exhibit, for example, non-zero linear AHE due to orbital magnetism through Berry curvature[33-35]. However, it turns out that all those altermagnets, showing linear AHE discussed so far, such as MnTe and RuO$_2$ thin films[35, 36] belong to ferromagnetic point groups.

**SYMMETRY OF DIAGONAL PIEZOMAGNETISM**

Piezoelectricity is the phenomenon of induced polarization, i.e., voltage gradient, with external stress. Similarly, piezomagnetism is the phenomena of inducing net magnetic moment with external stress, and there can exist Diagonal or Off-diagonal piezomagnetism. The experimental setup for Diagonal piezomagnetism along *z*, which is 1D, has broken {$T, M_x, M_y, R_x, R_y$} and unbroken {$1, I, M_z, R_z$}, and thus has broken {$I \otimes T, T, M_x, M_y, R_x, R_y$} with free rotation along *z*. MPGs for Diagonal piezomagnetism along *z*, requiring broken {$I \otimes T, T, M_x, M_y, R_x, R_y$} with free rotation along *z*, those for Diagonal piezomagnetism along *x*, requiring broken $\{I \otimes T, T, M_y, M_z, R_y, R_z\}$ with free rotation along *x*, and those for Diagonal piezomagnetism along *y*, requiring broken $\{I \otimes T, T, M_x, M_z, R_x, R_z\}$ with free rotation along *y*, are listed in Figs. 2, 3 and 4, respectively. For example, "broken $\{I \otimes T, T, M_x, M_y, R_x, R_y\}$ with free rotation along *z*" is a subset of "broken $\{I \otimes T, T, M_x, M_y, R_x, R_y, C_{3x}, C_{3y}\}$ with free rotation along *z*", which is the requirement for ferromagnetic point groups with magnetization along *z*. Therefore, all ferromagnetic point groups do exhibit Diagonal piezomagnetism along the magnetization direction. The only difference between the requirement for, for example, Diagonal piezomagnetism along *x* and that for ferromagnetic point groups with magnetization along *x* is broken $\{C_{3y}, C_{3z}\}$, and this difference is due to the breaking of both $C_{3y}$ and $C_{3z}$ by external stress along *x* in the case of Diagonal piezomagnetism along *x*. $C_3$ symmetry is defined to be along *z*, all MPGs, showing Diagonal piezomagnetism along *z*, belong to ferromagnetic point groups, but all MPGs with $C_{3z}$, showing Diagonal piezomagnetism along *x* or *y*, do not belong to ferromagnetic point groups, as shown in Fig. 3 and 4. Basically, breaking $C_{3z}$ by external stress is the essential part of the Diagonal piezomagnetism along *x* or *y* in MPGs with $C_{3z}$, which do not exhibit magnetization along *x* or *y* in zero strain.

**SYMMETRY OF CIRCULAR DICHROISM**

An experimental setup to measure Circular Dichroism (CD) is shown in Fig. 1(i) - (k). Circular Dichroism[37] includes the Faraday effect[38], natural optical activity, and Circular PhotoGalvanic Effect (CPGE)[39, 40]. Fig. 1(i) and (j) are linked through $\{M_x, M_y\}$, but each setup is invariant under $\{1, T \otimes R_x, T \otimes R_y\}$ and any spatial rotation around *z*. Thus, this setup has unbroken $\{T \otimes R_x, T \otimes R_y\}$ and broken $\{M_x, M_y\}$, so "broken $\{M_x, M_y\}$ + unbroken $\{T \otimes R_x, T \otimes R_y\} \otimes$ broken $\{M_x, M_y\}$" = broken $\{I \otimes T, M_x, M_y\}$ is required to have CD along *z* (CD$_z$). MPGs for CD$_z$, requiring broken $\{I \otimes T, M_x, M_y\}$, those for CD$_x$, requiring broken $\{I \otimes T, M_y, M_z\}$, and those for CD$_y$, requiring broken $\{I \otimes T, M_x, M_z\}$ are shown in Figs. 2, 3, and 4, respectively. Emphasize that for this symmetry consideration, we allow any spatial rotation around *z* freely. The broken-symmetry requirement for CD is a subset of those for Odd-order AHE, except that free rotations should be allowed for the symmetry consideration for CD. Thus, most of MPGs for non-zero Odd-order AHE, except $4'/m$, $4'm'm$, $\bar{4}'2'm$, $4'/mmm'$, $m\bar{3}$, $m\bar{3}m'$, $\bar{4}'$, $\bar{4}'2m'$, $\bar{4}'3m'$ for Odd-order AHE$_{yx}$, allow CD$_z$. In the case of all ferromagnetic point groups, which is a part of MPGs for Odd-order AHE, CD is always along the magnetization direction.

Now, CD can be nonreciprocal if all of {**T,R$_x$,R$_y$,C$_{3x}$,C$_{3y}$**} are broken additionally since any of {**T,R$_x$,R$_y$**} can link Fig. 1(i) and Fig. 1(k) in the condition of any free rotations along *z*. Thus, the requirement for Nonreciprocal CD (NCD) along *z* is broken {**I⊗T,T,M$_x$,M$_y$,R$_x$,R$_y$**} with free rotation along *z*, which is identical to the requirement for Diagonal piezomagnetism along *z*. Thus, all MPGs belonging to Ferromagnetic point group exhibit nonreciprocal CD. Thus, all MPGs for Diagonal piezomagnetism along one direction do exhibit NCD along the direction. All ferromagnetic point groups do exhibit NCD along the magnetization direction, which is associated with the Faraday effect. Note that breaking **C$_{3z}$** by external stress, discussed in the above Section, becomes also relevant for CD, since light propagation itself can also break **C$_{3z}$**. The following example illustrates how NCD can be linked with Diagonal piezomagnetism in nonferromagnetic point groups. Non-ferromagnetic MPGs, such as 32, 3*m* and $\bar{3}m$, should exhibit NCD$_x$ with zero NCD$_{y\text{ or }z}$, and they also exhibit Diagonal piezomagnetism along *x* (piezo*M$_x$*). In fact, Diagonal piezo*M$_x$* is the same with Diagonal even-order (e.g, 2$^{nd}$-order) current (light propagation or electric current)-induced *M*, and this induced *M* can result in NCD. Note that both space inversion and time reversal symmetries are broken in 32 and 3*m*, and their space-inversion domains do exhibit the identical NCD$_x$, but their time-reversal domains should display the opposite NCD$_x$. We emphasize that light-induced *M* can result in reciprocal CD as well as NCD. For example, reciprocal CD (i.e., natural optical activity) occurs in chiral point groups, and the light propagation in chiral materials does induce *M* along the light propagation direction in a linear fashion, which, in turn, results in *M*-induced CD[18]. Since the induced *M* flips its sign when light propagation direction flips by 180° (i.e. *M* is induced in a linear fashion), this CD effect is reciprocal. In other words, natural optical activity in chiral materials can be understood in terms of the Faraday effect due to light-induced *M*.

**SYMMETRY OF MAGNETO-OPTICAL KERR EFFECT (MOKE)**

The experimental setup to measure non-zero Magneto-Optical Kerr Effect (MOKE), i.e. the light-polarization rotation effect of reflected linearly-polarized light, is shown in Fig. 1(l)&(m). First, we note that this experimental setup is also invariant under any spatial rotations around *z*, so we can ignore or freely allow any rotations around *z* for symmetry considerations. The MOKE setup has broken {**T,M$_x$,M$_y$**}, so the requirement for MOKE along *z* (MOKE$_z$) is broken {**T,M$_x$,M$_y$**} with freely-allowed rotations around *z*. MPGs for MOKE$_z$, requiring broken {**T,M$_x$,M$_y$**} with free rotation around *z*, those for MOKE$_x$, requiring broken {**T,M$_y$,M$_z$**} with free rotation around $z$, and those for MOKE$_y$, requiring broken {**T**,**M$_x$**,**M$_z$**} with free rotation around $z$, are summarized in Figs. 2, 3 and 4, respectively. Note that, for example, in tetragonal point groups, **M$_x$** can be different from **M$_{xy}$**, and broken {**T**,**M$_x$**,**M$_y$**} with freely-allowed rotations around $z$ means broken {**T**,**M$_x$**,**M$_y$**,**M$_{xy}$**,**M$_{yx}$**} for those tetragonal point groups. Certainly, all ferromagnetic point groups have broken {**T**,**M$_x$**,**M$_y$**}, so they exhibit MOKE. It turns out that the requirement for diagonal linear magnetoelectric effects along $z$ is broken {**T**,**I**,**M$_x$**,**M$_y$**,**R$_x$**⊗**T**,**R$_y$**⊗**T**} with free rotation along $z$, which requires more broken symmetries than MOKE$_z$, so all linear magnetoelectrics along $z$ do exhibit MOKE$_z$. In fact, it turns out that all MPGs showing MOKE, but not belonging to ferromagnetic point groups, are diagonal linear magnetoelectrics. This new phenomena of MOKE in all linear magnetoelectrics can be considered as a result of magnetization induced by the presence of a surface in diagonal linear magnetoelectrics, since the presence of a surface is necessary for MOKE and any surface of all diagonal linear magnetoelectrics can have surface magnetization[41].

**CANDIDATE MATERIALS**

Using Figures 2-4, one can readily identify new materials suitable for measuring the various phenomena we have discussed. Materials belonging to any thirty-one ferromagnetic point groups (1, $\bar{1}$, 2, 2′, m, $m'$, 2/m, 2′/$m'$, 2′2′2, $m'm'm$, $m'm'$2, m′m2′, 4, $\bar{4}$, 4/m, 42′2′, 4$m'm'$, $\bar{4}$2′$m'$, 4/$mm'm'$, 3, $\bar{3}$, 32′, 3$m'$, $\bar{3}m'$, 6 , $\bar{6}$ , 6/m, 62′2′, 6$m'm'$, $\bar{6}m'$2′, 6/$mm'm'$) can exhibit all relevant phenomena. As we have discussed, these materials include seemingly-antiferromagnetic magnets such as Mn$_3$(Sn,Ge,Ga,Rh,Ir,Pt), as shown in Fig. 1(a), (b) and (d) and the so-called altermagnets such as MnTe and RuO$_2$ with the MPG $m'm'm$.[35, 36] Another examples are metallic cubic Pd$_3$Mn [42] and insulating NaMnFeF$_6$ [43] forming in ferromagnetic 32′ with unbroken {**C$_{3z}$**,**R$_x$**⊗**T**}. The 32' point group, allowing all of these phenomena, e.g. magnetization$_z$, Odd-order AHE$_{yx}$, Odd-order AHE$_{xy}$, Odd-order AHE$_{zx}$, NCD$_{y\,or\,z}$, Diagonal piezomagnetism$_{y\,or\,z}$, and MOKE$_{y\,or\,z}$, so Pd$_3$Mn and NaMnFeF$_6$ can be studied for these phenomena. Note that since magnetization of 32' is along z, so Odd-order AHE$_{yx}$, Odd-order AHE$_{xy}$, NCD$_z$ , Diagonal piezomagnetism$_z$ and MOKE$_z$ are entirely expected; however, they can also exhibit Odd-order AHE$_{zx}$, NCD$_y$, Diagonal piezomagnetism$_y$ and MOKE$_y$, even though magnetization along $y$ is zero. These off-magnetization-direction phenomena have never been observed.

Emphasize that the current for AHE can be electric current or other propagating quasiparticles such as thermal current, propagating lights, magnons and phonons. As discussed earlier, Odd-order AHE, associated with symmetric tensors, can occur in true antiferromagnets in the MPGs of 4', $\bar{4}$', 4'/$m$, 4'2'2 (or 4'22'), 4'$m$'$m$ (or 4'$mm$'), $\bar{4}$'2'$m$, $\bar{4}$'$m$'2 ($\bar{4}$'2$m$'), 4'/$mmm$'. MPG of the magnetic ground state of $RuO_2$ in bulk form is 4'/$mmm$',[44] which does not belong to the ferromagnetic point group. Thus, high-odd-order AHE can be expected in $RuO_2$, which requires future experimental confirmation. The magnetic states of insulating $Er_2Ge_2O_7$ and $Pb(TiO)Cu_4(PO_4)_4$ are non-ferromagnetic 4'22' with unbroken {$C_{4z} \otimes T, R_z, R_x, R_y, R_{xy} \otimes T, R_{yx} \otimes T$}, allowing Odd-order $AHE_{yx,xy}$, $CD_{x, y, xy, yx\ or\ z}$, and $MOKE_{x, y, xy\ or\ yx}$. Thus, it is imperative to measure Odd-order $AHE_{yx,xy}$, $CD_{x, y, xy, yx\ or\ z}$, and $MOKE_{x, y, xy\ or\ yx}$ in $Pb(TiO)Cu_4(PO_4)_4$ and $Er_2Ge_2O_7$. Note that this Odd-order $AHE_{yx,xy}$ can be observed as an anomalous thermal Hall effect in those insulating 4'22' systems without any ferromagnetic moment. The MPG of the magnetic state of $Fe_2Mo_3O_8$ is 6'$mm$',[45] which can exhibit Off-diagonal piezomagnetism$_{xy}$, Odd-order $AHE_{zy}$, Diagonal piezomagnetism$_x$, $NCD_x$, and $MOKE_x$ (see Fig. 2), which need to be experimentally verified. Note that when we have Diagonal odd-order (such as linear) piezomagnetism$_x$ and Off-diagonal odd-order (such as linear) piezomagnetism$_{xy}$, applying electric field or current along $x$ or $y$ will induce magnetization along $x$, which is an even-order (such as quadratic) with applied electric field/current, so the sign change of applied electric field/current will not change the sign of induced magnetization. Finally, note that the MPG of $CsFeCl_3$ below $T_N$=4.7 K is $\bar{6}$'$m$2', while the point group of it above $T_N$ is centrosymmetric 6/mmm. $\bar{6}$'$m$2' has unbroken {$C_{6z} \otimes I \otimes T, M_x, M_{xy}, M_y, M_z \otimes T, C_{3z}$}, so can exhibit Odd-order $AHE_{zy}$, $NCD_x$, Diagonal piezomagnetism$_x$, and $MOKE_x$. Thus, insulating $CsFeCl_3$ in a non-ferromagnetic point group with zero magnetization can exhibit NCD such as the Faraday effect, which has been always thought to be confined in ferromagnetic systems.

**CONCLUSION**

Our SOS concept incorporates the symmetry relationship between specimen and experimental setup, encompassing measurables and sample environment, without considering local coupling or relevant tensorial terms. The SOS approach can tell if a certain measurable relevant to a particular phenomenon is zero, non-zero odd-order, or non-zero even-order. By employing this SOS approach, we have successfully identified all MPGs relevant for each of ferromagnetism-like phenomena, including magnetic attraction/repulsion, diagonal piezomagnetism, nonreciprocal circular dichroism (such as Faraday effect), odd-order (including linear) anomalous Hall effect, and magneto-optical Kerr effect. The ferromagnetism-like phenomena can manifest only in two ways: first, through non-zero magnetization in ferromagnetic point groups, where symmetry permits non-zero magnetization; and second, through magnetization induced by external perturbations such as electric current flow, electric fields, light propagation, or strain. Undoubtedly, the categorized MPGs for each ferromagnetism-like phenomenon, along with our SOS approach, will serve as crucial guidance for future advancements in magnetism-related science and technology.

**ACKNOLEDGEMENT:** The work was supported by the center for Quantum Materials Synthesis (cQMS), funded by the Gordon and Betty Moore Foundation's EPiQS initiative through grant GBMF10104, and by Rutgers University

**AUTHOR CONTRIBUTIONS:** S.W.C. conceived and supervised the project. F.-T.H. conducted magnetic point group analysis. S.W.C. wrote the remaining part.

**COMPETING INTERESTS:** The authors declare no competing interests.

**DATA AVAILABILITY:** All study data are included in the article.

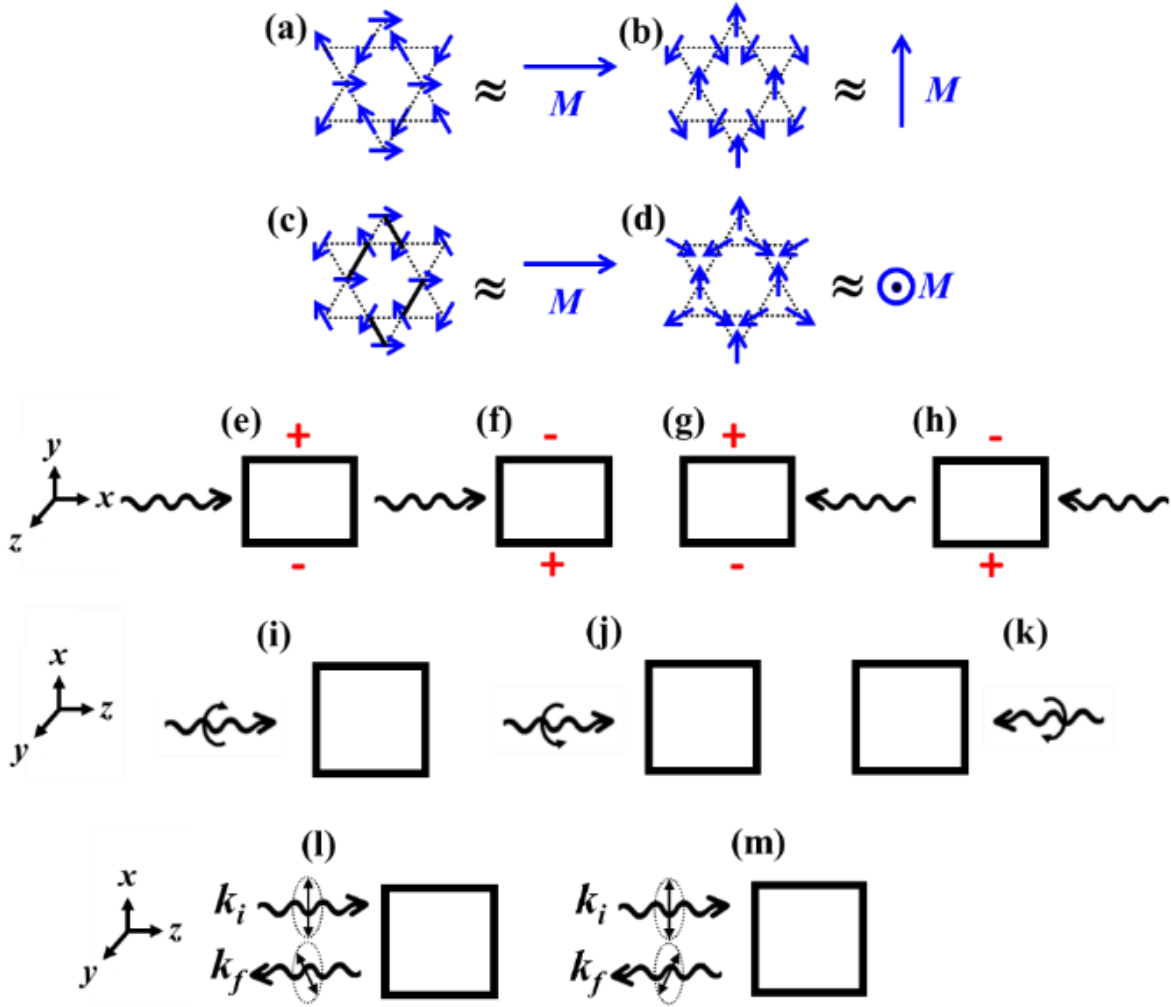

**Fig. 1 Various specimens and experimental setups for AHE, circular dichroism and MOKE.** (a)-(d) are for Mn$_3$Sn (*Cmc'm'*), Mn$_3$(Ge,Ga) (*Cm'cm'*), kagome lattice with lattice distortions, shown with solid-line bonds (*P2$_1$/n*), and Mn$_3$(Rh,Ir,Pt) ($R\bar{3}m'$), respectively. (e)-(h) Four experimental setups to measure AHE, i.e. transverse voltage induced by current. (i)-(k) Three experimental setups to measure circular dichroism. (l)- & (m) Two experimental setups to measure MOKE.

**Odd-order**
**AHE$_{yx}$/AHE$_{yx,xy}$**
$4'/m, 4'm'm, \bar{4}'2'm$
$4'/mmm', m\bar{3}, m\bar{3}m'$

$\bar{4}', \bar{4}'2m', \bar{4}'3m'$           $4', 4'22', 4'32'$

23

**FM$_z$ (1D)**

**MOKE$_z$ (1D)**    NCD$_z$    $1, \bar{1}, 2', m'$          **CD$_z$ (1D)**

$\bar{1}', 2/m'$    Diagonal   $2'/m', 2'2'2$
$m'm'm'$    piezoM$_z$   $m'm'm, m'm'2$      $11', 21', 2221'$
$\bar{3}', \bar{3}'m'$    (1D)    $4, \bar{4}, 4/m, 42'2'$      $31', 321'$
$4/m', 4/m'm'm'$      $4m'm', \bar{4}2'm', 4/mm'm'$   $41', 4221'$
$\bar{6}', 6/m', \bar{6}'m'2,$      $3, \bar{3}, 32', \bar{3}m', 3m'$      $61', 6221'$
$6/m'm'm'$      $6, \bar{6}, 6/m, 62'2', 6m'm'$    $4321', 231'$
$m'\bar{3}', m'\bar{3}'m'$      $\bar{6}m'2', 6/mm'm'$

$2, 222, 32, 422$
$622, 432$

**Fig. 2 Magnetic point groups for various ferromagnetism-like phenomena along *z*.** Odd-order AHE$_{yx}$ is expected to be identical to Odd-order AHE$_{xy}$, except for the possible sign difference. Green: MPGs of diagonal linear magnetoelectric; Blue: MPGs of chiral point group; Turquoise: MPGs of linear magnetoelectric and chirality.

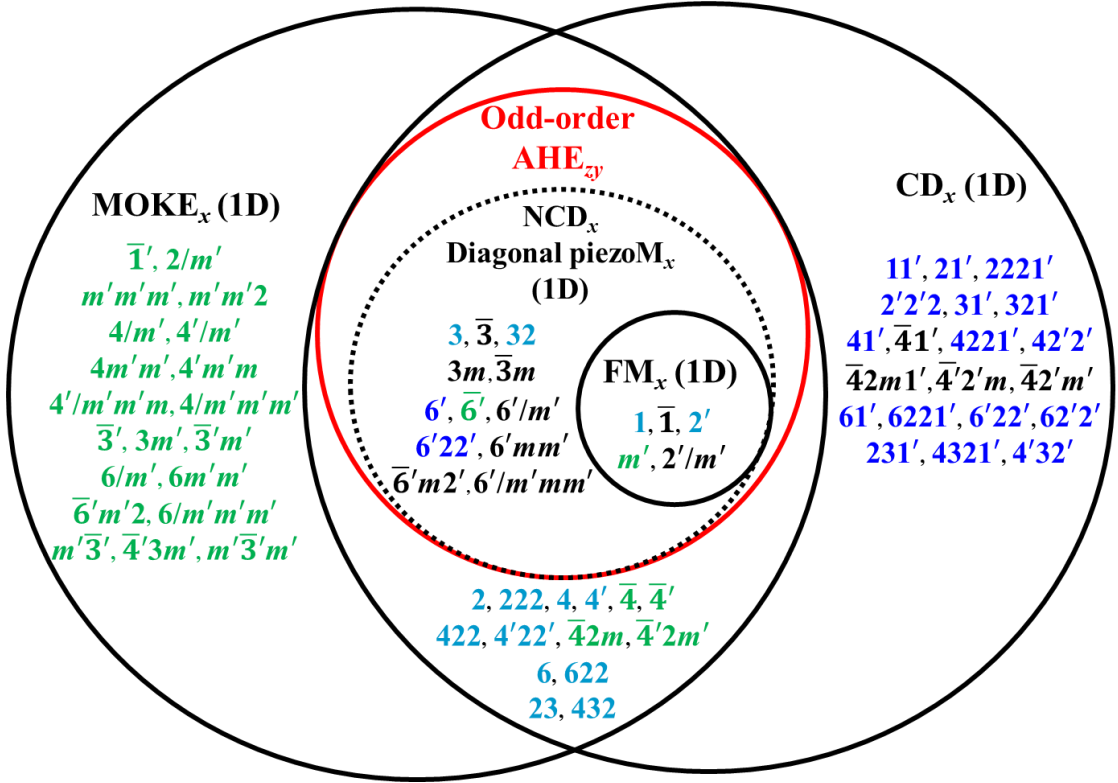

**Fig. 3 Magnetic point groups for various ferromagnetism-like phenomena along *x*.** Only in ferromagnetic point groups, Odd-order AHE$_{zy}$ is expected to be identical to Odd-order AHE$_{yz}$, except the possible sign difference. Green: MPGs of diagonal linear magnetoelectric; Blue: MPGs of chiral point group; Turquoise: MPGs of both.

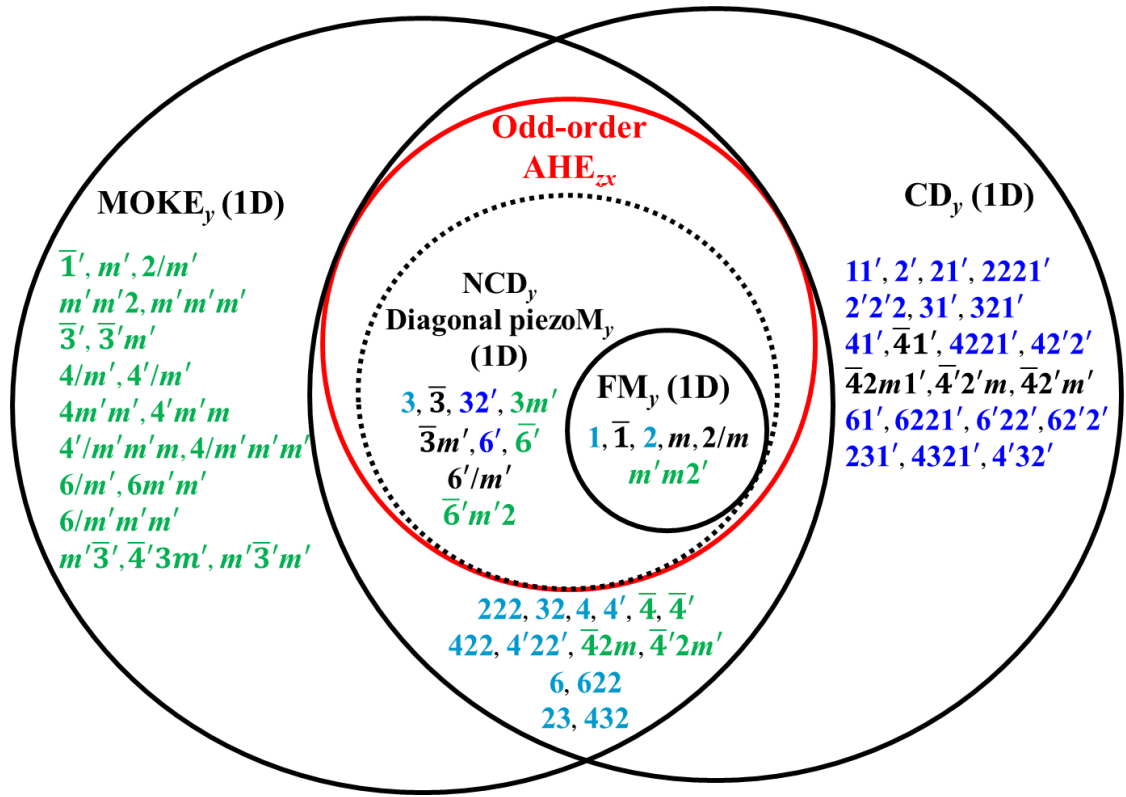

**Fig. 4 Magnetic point groups for various ferromagnetism-like phenomena along *y*.** Only in ferromagnetic point groups, Odd-order AHE$_{zx}$ is expected to be identical to Odd-order AHE$_{xz}$, except the possible sign difference. Green: MPGs of diagonal linear magnetoelectric; Blue: MPGs of chiral point group; Turquoise: MPGs of both.

**Table 1, Required broken symmetries and relevant MPGs for certain measurables or phenomena.** The subscripts of $x$, $y$ and $z$ mean the relevant orientations: for example, Odd-order AHE$_{yx}$ means current along $x$ and Hall voltage along $y$, and Diagonal piezomagnetism$_z$, Circular Dichroism$_z$, and MOKE$_z$ are along $z$. Note that Odd-order AHE$_{yx}$ = Odd-order AHE$_{xy}$ except sign in all ferromagnetic point groups. We emphasize that when we consider broken symmetries, we ignore or freely allow any rotations around the 1D direction for 1D phenomena such as Diagonal piezomagnetism$_z$, Circular Dichroism$_z$, and MOKE$_z$ (FR$_z$ means free rotation around $z$). For example, MPG $\bar{4}1'$ has unbroken **T** and broken **I**, but **I**⊗**C$_{4z}$** is unbroken, so **I**⊗**T** is unbroken with FR$_z$ and it cannot exhibit Circular Dichroism$_z$; however, $\bar{4}1'$ can exhibit Circular Dichroism$_{\text{x or y}}$.

| Measurables | Required broken symmetries | Relevant MPGs |
| --- | --- | --- |
| Non-zero magnetization$_z$ (FR$_z$) | {**I**⊗**T**,**T**,**M$_x$**,**M$_y$**,**R$_x$**,**R$_y$**,**C$_{3x}$**,**C$_{3y}$**} | All ferromagnetic point groups |
| Diagonal piezomagnetism$_z$ (Nonreciprocal optical activity) (FR$_z$) | {**I**⊗**T**,**T**,**M$_x$**,**M$_y$**,**R$_x$**,**R$_y$**} | All ferromagnetic point groups |
| Odd-order AHE$_{yx}$ | {**I**⊗**T**,**T**,**M$_x$**,**M$_y$**,**R$_x$**,**R$_y$**,**C$_{3x}$**,**M$_z$**⊗**T**,**R$_z$**⊗**T**} | All ferromagnetic point groups & 4', $\bar{4}$', 4'/$m$, 4'$m'm$, $\bar{4}$'2'$m$. ($\bar{4}$'2$m'$, 4'22', 4'/$mmm'$, 23, $m\bar{3}$, 4'32', $\bar{4}$'3$m'$, $m\bar{3}m'$ for current along $xy$ or $yx$) |
| Circular Dichroism$_z$ (FR$_z$) | {**I**⊗**T**,**M$_x$**,**M$_y$**} | All ferromagnetic point groups & chiral point groups |
| MOKE$_z$ (FR$_z$) | {**T**,**M$_x$**,**M$_y$**} | All ferromagnetic point groups & All diagonal linear magnetoelectrics (broken {**T**,**I**,**M$_x$**,**M$_y$**,**R$_x$**⊗**T**,**R$_y$**⊗**T**} with FR$_z$) |